\documentclass{svjour3}                     
\smartqed  
\usepackage[normalem]{ulem}
\usepackage{graphicx}
\usepackage{color}
\usepackage{amsmath}
\usepackage{ulem}

\usepackage{color}

 \journalname{Scientometrics}
\begin{document}

\title{
Predicting Results of the Research Excellence Framework using Departmental $h$-Index -- Revisited
 %
%
}


\author{O.~Mryglod \and R.~Kenna \and Yu.~Holovatch \and B.~Berche }

\institute{O. Mryglod \at
              Institute for Condensed Matter Physics of the National Academy of Sciences of Ukraine,
              1 Svientsitskii Str., 79011 Lviv, Ukraine\\
              \email{olesya@icmp.lviv.ua}
           \and
            R. Kenna \at
              Applied Mathematics Research Centre, Coventry University,
              Coventry, CV1 5FB, England
           \and
            Yu. Holovatch \at
              Institute for Condensed Matter Physics of the National Academy of Sciences of Ukraine,
              1 Svientsitskii Str., 79011 Lviv, Ukraine
           \and
            B. Berche \at
               Universit\'e de Lorraine, Statistical Physics Group, IJL, UMR CNRS 7198, Campus de Nancy, B.P. 70239, 54506 Vand\oe uvre l\`es Nancy Cedex, France Vand\oe uvre l\`es Nancy Cedex, France
}

\date{Received:  / Accepted: date}

\maketitle

\begin{abstract}
We revisit our recent study [{\emph{Predicting results of the Research Excellence Framework using departmental h-index}}, Scientometrics, 2014, 1-16; arXiv:1411.1996] in which we attempted to predict outcomes of the UK's Research Excellence Framework (REF~2014) using the so-called departmental $h$-index.
Here we report that our predictions failed to anticipate with any accuracy either overall REF outcomes or movements of individual institutions in the rankings relative to their positions in the previous Research Assessment Exercise (RAE~2008).
\end{abstract}

\section{Introduction}

The results of the last national exercise in research assessment in the UK -- the {\emph{Research Excellence Framework}} (REF) -- became available at the end of December 2014.
As with its precursor -- the {\emph{Research Assessment Exercise}} (RAE) -- a vast amount of discussion and debate surrounds the process, especially concerning the merits of such peer-review based exercise themselves, but also about whether or not they could sensibly be replaced or supplemented by the usage of quantitative metrics.

In this context, and before the results of the REF were announced, we attempted to use departmental $h$-indices to predict REF results \cite{2014_Scientometrics}.
In particular, after demonstrating that the  $h$-index is better correlated with the  results from RAE~2008 than a competing metric, we determined $h$ values for the period examined in the REF.
We looked at four subject areas:  biology,  chemistry,  physics and sociology and  placed our predictions in the public domain, including in this journal, promising to revisit the paper after REF results were announced \cite{2014_Scientometrics}.

Here we fulfill that promise and compare $h$-predictions with the outcomes of the REF: We report that our predictions were wildly inaccurate.

Our previous paper drew considerable interest in the media and on the blogosphere.
Anticipating a similar degree of interest in the results of our analysis presented here,
we also reflect on  its implications.

\section{Predicting the REF}

The results of both RAE~2008 and REF~2014 are disseminated as quality profiles which partition submitted research of  higher education institutes (HEI's) into five  bands, decreasing in quality from world-leading to unclassified. To capture this gradation in a single summary statistic, in \cite{2014_Scientometrics} we used a post-RAE funding formula devised by the {\emph{Higher Education Funding Council for England}}  and denoted by $s$.
We observed  discipline dependent correlations between  $s$ values from the RAE~2008 and departmental $h$-indices: The  correlation coefficients varied between 0.55 and 0.8.
Such results are not good enough to consider replacing RAE/REF by citation-based measures since even small differentials in  rating can have considerable consequences for HEI's in terms of  reputation and the state funding  received.
Nevertheless, we considered it interesting to check the extent to which the results from  REF~2014 would correlate with departmental $h$-indices and whether predictions could be made.

Following the notation of \cite{2014_Scientometrics}, departmental Hirsch indices based on the RAE~2008 assessment period from 2001 to 2007 are denoted $h_{2008}$.
Those for papers published between 2008 and 2013 (the REF~2014 assessment period) are denoted ${h}_{2014}$ here.
We are interested in two types of prediction, each measured by correlation coefficients.
Firstly, a ``global'' picture,
representing comparisons between $s$-values delivered by the REF and ${h}$-values delivered by Hirsh indices,
is gauged by Pearson's coefficient $r$ and Spearman's rank correlation coefficients $\rho$.
Individual universities  are also interested in a more local picture - whether they move up or down in the REF rankings relative to HEI's competing in a particular subject area.
It is not unreasonable to assume that a shift upwards or downwards in the Hirsch-index rankings would be accompanied by movement in the same direction in the RAE/REF rankings.
In this manner, one may seek to predict, not the exact positions of various institutions in the rankings, but relative  directions of movement.
These are also measured by correlation coefficients.

Therefore, we seek to address two questions: what is the correlation between REF~\-2014 scores and the corresponding departmental $h$-indices and is it possible to predict the tendencies of changes in the rankings of submitting institutes.

\section{Results}

Before delivering the results of our analysis, we comment that the list of submitters to REF~2014 is different to that of RAE~2008.
Moreover, due to technical reasons it was not possible to obtain the citation data, and therefore to calculate $h$-indices, for a small number of institutions (those that were not listed in  the {\emph{Scopus}} database used after refining the search results).
We have to limit our analysis to those HEI's for which  RAE, REF and $h$-index scores are available.
This reduction in data-set size can affect correlation coefficients.
For example, the correlation coefficients $r\approx0.74$ and $\rho\approx 0.78$ between RAE~2008 and $h_{2008}$, published previously for Biology (see \cite{2014_Scientometrics}, table 1), were calculated for the 39 groups which submitted to RAE~2008 and for which the Scopus data were available.
But if we drop the 8 HEI's which  did not submit in this unit of assessment in REF~2014, the resulting correlation coefficients values change to $r\approx 0.55$ and $\rho\approx 0.61$, respectively.
This caveat notwithstanding we compare the ranked lists of HEI's for which all four scores are available. These comprise 31 Biology groups, 29 Chemistry groups, 32 Physics groups and 25 Sociology groups.

We also remind that both the RAE and the REF had three components, one of which involved outputs (publications) only. (The other components, which contributed less than outputs to overall quality profiles, were research environment and esteem or non-academic impact.) Since the Hirsch index is a function of citations to publications, we compare both to overall RAE/REF scores and to the scores coming from  outputs only.

The calculated correlation coefficients are given in table~\ref{tab1}.
In the table we present separately the results for overall $s$ values (upper part) and for $s$-values corresponding to outputs only (lower part).
Comparing the values presented in the columns 2--5, one can see that the RAE~2014 scores are not much better correlated with departmental $h$-indices than RAE~2008.
The correlation coefficients are positive but still not strong enough to make accurate predictions or to replace REF with metrics.
As already found in \cite{2013_Scientometrics}, the output component of REF is  more weakly correlated with the citations-based metric for Biology, Chemistry and Physics.

\begin{table}[b]
\caption{The values of Pearson coefficients $r$ and Spearman rank correlation coefficients $\rho$, calculated for different disciplines for different pairs of measures. The numbers of HEI's which were taken into account to calculate the corresponding pair of coefficients (Pearson and Spearman) are given in  parentheses. All values, except those in boldface, are statistically significant at the level $\alpha=0.05$.
The upper part uses $s$ values from the overall RAE and  REF results ($s_{\rm{RAE}}$ and $s_{\rm{REF}}$, respectively) while the lower part corresponds to the results for outputs only.
Correlations between predicted and actual directions of shift (up or down) in the ranked lists based are given in the final columns.}
\begin{center}
\begin{tabular}{|l||l|l|l|l|l|l||l|}
\hline
 OVERALL \hspace{0.8cm} & \multicolumn{2}{|c|}{$s_{\mathrm{RAE}}$ vs. $h_{2008}$}& \multicolumn{2}{|c|}{$s_{\mathrm{REF}}$ vs. ${h}_{2014}$}&
$\uparrow \downarrow$\\
\hline  & $r$ & $\rho$& $r$ & $\rho$&$r$\\
\hline Biology (31) &0.55&0.61&0.58&0.63&\textbf{--0.15}\\
\hline Chemistry (29) & 0.80&0.83&0.84&0.89&\textbf{0.05}\\
\hline Physics (32) &0.49&0.55&0.55&0.50&\textbf{0.26}\\
\hline Sociology (25) &0.50&\textbf{0.39}&0.59&0.60&\textbf{0.18}\\
\hline
\end{tabular} \label{tab1}
\vspace{0.5cm}
\begin{tabular}{|l||l|l|l|l|l|l||l|}
\hline
OUTPUTS ONLY & \multicolumn{2}{|c|}{$s_{\mathrm{RAE}}$ vs. $h_{2008}$}& \multicolumn{2}{|c|}{$s_{\mathrm{REF}}$ vs. ${h}_{2014}$}&
$\uparrow \downarrow$\\
\hline  & $r$ & $\rho$& $r$ & $\rho$&$r$\\
\hline Biology (31) &0.44&0.51&0.40&0.42&\textbf{--0.33}\\
\hline Chemistry (29) & 0.74&0.71&0.71&0.72&\textbf{0.20}\\
\hline Physics (32) &0.44&0.51&0.39&\textbf{0.36}&\textbf{0.02}\\
\hline Sociology (25) &0.41&\textbf{0.29}&0.71&0.68&\textbf{0.06}\\
\hline
\end{tabular} \label{tab1}
\end{center}
\end{table}

The last column in the table~\ref{tab1} indicates the correlations between predicted and actual directions of shift (up or down) in the ranked lists based on different measures.
The correlations are weakly positive or even negative.
This approach, however, does not take into account different magnitudes of $h$-index shifts. We surmised that there may be a threshold such that only $h$-index changes greater than a critical value tend to manifest changes in the same direction in the $s$-ranks.
However  no such threshold was found at least using the limited data available.
These mean that it is not possible to use the departmental $h$-index in this manner to predict whether a given  HEI will move up or down in the REF  rankings relative to other HEI's.

\section{Conclusions and Discussion}

Here we present conclusions  coming from the two parts of this study \cite{2014_Scientometrics}. Given the broad levels of  interest in the first part \cite{2014_Scientometrics}, we feel an extended discussion on the role of metrics in national research assessment exercises of the types considered here is warranted.

It is well documented that the REF itself is flawed (lack of robust calibration process, insufficient scrutiny of each and every sub-discipline, inevitable human error and bias, lack of robust normalisation between disciplines, etc.) and Goodhart's law informs us that when measures becomes targets, they cease to be good measures.
Despite these shortcomings, peer review is currently widely considered to be the most acceptable way to evaluate research quality.
It is the only national process in current usage in the UK is the REF, so any replacement would presumably have to be able to mimic it accurately to be accepted by policy makers and the academic community.

We have investigated whether departmental  $h$-indices could play such a role.
We found that the correlations between departmental $h$-scores  and REF~2014 results are more or less the same as those between the former and RAE~2008 \cite{2014_Scientometrics}.
Although they sometimes correlate well, the similarities are not good enough to make accurate predictions; $h$-indices used in this way do not track the peer review exercises well enough for them to form an component of, or substitute for those exercises. Additionally, we found very poor correlations between the predicted and actual changes in the ratings. This means that the departmental $h$-index does not offer a way to foretell the direction of changes of universities in the rankings in these subject areas.

It is worthwhile taking a step back to review what we are attempting to do with scientometrics in the context of national assessment exercises.
Academic research is a special kind of activity, often founded purely on curiosity.
Although applications may not be obvious in the short term, curiosity-driven research has led to some of the most important practical advances our civilisation has produced, including discoveries in medicine and technology.
Some of these advances have arisen decades after the scientific discoveries which underlie them.
Since commercial exploitability may be impossible to predict in a reasonable time frame or entirely absent from blue-skies research, curiosity-driven research is mostly carried out at universities and is funded by the public purse.
The REF, and its precursor, the RAE, are intended to monitor this public investment in the UK.
Other countries have other schemes, many involving the use of metrics.
The  pertinent question is whether these are fit for purpose.

Belief that  metrics in current use are counter-productive  has led to a recent ground\-swell of opinion against them -- see, e.g., the San Francisco Declaration on Research Assessment \cite{DORA}.
In France the CNRS has questioned the use of bibliometric indicators --- including the $h$-index --- in the evaluation of scientific  research~\cite{CNRS}.
One argument is that, in the increasingly managed  environments of many universities around the world,
where academic freedom has already lost ground to semi-industrialised processes,
the introduction of  metrics would further undermine environments for basic research.
In seeking to maximise metric scores, fashionable and incremental research may be promoted over foundational scientific inquiry.
This is potentially devastating to an endeavour which is at the very heart of what it
is to be human and a foundation of our society -- curiosity itself.

So is there a place for metrics in future national assessment exercises?
As for any other blunt tool, quantitative metrics can be useful if used in the correct manner, by informed subject experts.
But in the wrong hands, they can be dangerous.
Our study shows that a very different landscape would have emerged in the UK if REF~2014 had been entirely and simplistically based on the automated departmental $h$-index.
A wise academic subject expert can, perhaps, use such a metric to gain perspective in combination with other approaches, taking into account many nuances such as scientific context, subject history and history of science generally, technical aspects, future perspectives, interdisciplinarity and so on.
Clearly, however, over-reliance on a single metric by persons who are not subject experts could be misleading, especially in increasingly managed landscapes in which academic traditions are  diminished or eroded.


\section*{Acknowledgements}
This work was supported by the 7th FP, IRSES project No. 269139 ``Dynamics and cooperative phenomena in complex physical and biological environments'', IRSES project No. 295302 ``Statistical physics in diverse realizations'' and IRSES project No. 612707 ``Dynamics of and in Complex Systems''.

 
\end{document}